\documentclass[12pt,preprint]{aastex}

\slugcomment{2015 September 05}

\shorttitle{Asteroid 2003 EH$_{1}$}
\shortauthors{Kasuga et al.}

\begin{document}


\title{Physical Observations of (196256) 2003 EH$_{1}$, Presumed Parent of the Quadrantid Meteoroid Stream}


\author{Toshihiro Kasuga\altaffilmark{1,2}
and 
            David Jewitt\altaffilmark{2,3}
            } 	
\email{kasuga@perc.it-chiba.ac.jp}

%\author{R. J. Hanisch\altaffilmark{5}}
%\affil{Space Telescope Science Institute, Baltimore, MD 21218}


\altaffiltext{1}{Planetary Exploration Research Center, Chiba Institute of Technology, 2-17-1 Tsudanuma, Narashino, Chiba 275-0016, Japan}
\altaffiltext{2}{Visiting Astronomer, Kitt Peak National Observatory, National Optical Astronomy Observatory, 
which is operated by the Association of Universities for Research in Astronomy (AURA) 
under cooperative agreement with the National Science Foundation.}
\altaffiltext{3}{Department of Earth, Planetary and Space Sciences and Department of Physics and Astronomy, University of California at Los Angeles, 595 Charles Young Drive East, Los Angeles, CA 90095-1567, USA}

%\altaffiltext{2}{Society of Fellows, Harvard University.}
%\altaffiltext{3}{present address: Center for Astrophysics,
%    60 Garden Street, Cambridge, MA 02138}
%\altaffiltext{4}{Visiting Programmer, Space Telescope Science Institute}
%\altaffiltext{5}{Patron, Alonso's Bar and Grill}


\begin{abstract}

The near-Earth asteroid (196256) 2003 EH1 has been suggested to have a dynamical association 
with the Quadrantid meteoroid stream.  We present photometric 
observations taken to investigate the physical character of this body
and to explore its possible relation to the stream.  We find no evidence for on-going mass-loss.  A model fitted to the
 point-like surface brightness profile at 2.1 AU limits the fractional contribution to the integrated brightness by near-nucleus coma  to $\leq$ 2.5 \%.  
Assuming an albedo equal to those typical of cometary nuclei ($\it p_{\rm R}$=0.04), 
we find that the effective nucleus radius is $r_e$ = 2.0$\pm$0.2\,km. 
Time-resolved ${\it R}$-band photometry can be fitted by a two-peaked lightcurve having a
rotational period of 12.650$\pm$0.033\,hr. The range of the lightcurve, 
$\Delta  m_{\rm R}$= 0.44 $\pm$ 0.01\, mag, is indicative of an elongated 
shape having an axis ratio $\sim$1.5 projected into the plane of the sky.  
The asteroid shows colors slightly redder than the Sun,  
being comparable with those of C-type asteroids.     
The limit to the mass loss rate set by the absence of resolved coma is $\lesssim$ 2.5$\times$ 10$^{-2}$ kg ${\rm s^{-1}}$, corresponding 
to an upper limit on the fraction of the surface that could be sublimating water ice $f_A$ $\lesssim$ 10$^{-4}$.  
Even if sustained over the 200-500 yr dynamical age of the Quadrantid stream, 
the total mass loss from 2003 EH1  would be too small
to supply the reported stream mass ($10^{13}$\,kg),
implying either that the stream has another parent or that mass loss from 2003 EH1 is episodic.   

\end{abstract}





\section{Introduction}

The near-Earth Asteroid (196256) 2003 EH1 (hereafter 2003 EH1) was discovered on UT 2003 March 6 
in the course of the Lowell Observatory Near-Earth-Object Search (LONEOS) \citep{S03}.  
Dynamical studies show that the asteroid is associated with, and is presumed to be the parent body of,
the Quadrantid meteoroid stream \citep{Jennis04,Wi04,WB05,Ba08,Jo11,Abedin15}.
The orbit has semimajor axis ${\it a}$ = 3.126\,AU, eccentricity ${\it e}$= 0.619, 
and inclination ${\it i}$=$70^{\circ}.8$ (NASA/JPL HORIZON).  
The Tisserand parameter with respect to Jupiter, ${\it T_{\rm J}}$ = 2.063, is consistent with the 
 dynamical classification of 2003 EH1 as a Jupiter-family comet (JFC), although no activity has yet been reported.      
A straightforward interpretation is that 2003 EH1 is a 
dormant or weakly active comet \citep{Koten06,Ba08,Bo10,Tanc14}.

Dynamical studies of the recent ($<$10$^4$ yr) evolution of the orbit of 2003 EH1 under the action of planetary perturbations are suggestive in this regard.   
The semimajor axis lies close to the 2:1 mean-motion resonance with Jupiter at 3.27 AU, causing strong orbital variations that drive 2003 EH1 
into a sun-approaching dynamical state  \citep{WB05,Nes13b,FJ14}.   
Numerical integrations show that the perihelion distance has increased approximately linearly with time from 0.2 AU 1000 years ago to the present-day value of 1.2 AU.  The minimum $q$ $\sim$ 0.12\,AU (${\it e}$ $\sim$ 0.96) 
occurred only $\sim$1500\,yr ago  \citep{Nes13b,FJ14}.  
As a result, it is reasonable to expect that the surface layers should have been devolatilized at the high temperatures reached near past perihelia, leading to the present, apparently inert state.

The Quadrantid meteor shower was first reported in 1835 \citep{Q39}.   
The shower has a very short duration in its core activity (Earth crosses the core stream in $\sim$0.5 day)
superimposed on a broader, long-lived background activity (crossing time $\sim$4 days), suggesting that 
young and old meteoroid streams coexist \citep[][and references therein]{WB05}.  The width of a meteor stream increases with age, as a result of the progressive influence of planetary perturbations. The small width of the Quadrantid core stream indicates ejection ages only $\sim$200-500 years  \citep{Jennis04,Wi04,WB05,Abedin15} and there is some suggestion that the first reports of meteoroid stream activity coincide with the formation of the stream.  On the other hand, the broader  background stream implies larger ages of perhaps $\sim$3,500 years or more 
\citep{Oh95,WB05,KN07,Oh08}. 
 Comet 96P/Machholz is also suspected to form part of the ``Quadrantid complex'', possibly releasing meteoroids 
between 2,000--5,000 years ago~\citep{McI90, BaO92, Go92, JJ93,WB05}.   
Comet 96P/Machholz currently has a small perihelion orbit 
($\it a$= 3.034\,AU, $\it e$= 0.959,  $\it i$= $58^{\circ}.312$ and 
$\it q$= 0.124\,AU from NASA/JPL HORIZON) substantially different from that of 2003 EH1.
Despite these differences, the rapid dynamical evolution shows that it is possible that 2003 EH1 is a split fragment 
of  96P/Machholz or that both were released from a now defunct  
 precursor body~\citep[together defining the Machholz complex:][]{SC05}.  
One or both of these bodies  can be the parents of 
the Quadrantid meteoroids~\citep{KN07,Ba08,Nes13a,Nes13b,Nes14}.

The  small lifetime of the Quadrantid stream suggests that 2003 EH1 could still be active, particularly when in the small-perihelion orbital state.  
In this paper we report the first measurements of the physical properties of 2003 EH1, including colors, 
limits on coma activity, size, mass loss rate, fractional active area on the object 
and rotational period and further discuss the possible relation of this body to the Quadrantid stream and complex.

% --------------------------------------------------------------------------------------

\section{Observations}

We observed on the nights of UT 2013 August 8, 9 and 12 
using the Kitt Peak National Observatory 2.1\,m diameter telescope (hereafter, KPNO\,2.1) 
in Arizona and on  October 2 at the Keck-I 10\,m diameter telescope at the top of Mauna Kea, Hawaii.  
The KPNO\,2.1 employed a STA3 4000 $\times$ 2600 pixel charged-coupled device (CCD) 
camera at the f/7.5 Cassegrain focus.   
We used a 2$\times$2 binned image scale $0\arcsec.298$ ${\rm pixel}^{-1}$, giving  
a field of view (FOV) approximately $9^{'}.6\times$6$^{'}$.7.
On Keck-I,  
the Low Resolution Imaging Spectrometer (LRIS) camera \citep{Oke1995} was used to image the object. 
The LRIS camera has two separate channels having red and blue optimized CCD imagers separated by a dichroic filter. 
One is a red-side detector having a mosaic of two LBNL 2000 $\times$ 4000 pixels \citep{Roc10}
and the other is a blue-side detector having a mosaic of two  2K$\times$4K Marconi CCDs, both 
with imaging scale 0.$^{''}$135 ${\rm pixel}^{-1}$. 
The field of view in both modes of operation is $6^{'}.0 \times$ 7$^{'}$.8.
For imaging data, both telescopes were tracked non-sidereally to follow the motion of 2003 EH1.   
On KPNO\,2.1, images were taken through the Johnson-Kron-Cousins $\textit{BVRI}$-filter system.    
On Keck-I, images in the $\it R$-filter were recorded using the red-side detector of LRIS. 
The images were flattened by subtracting a bias image and dividing by a bias-subtracted flat-field image 
 constructed using artificial illumination of the inside the each dome for each filter.   
Photometric calibrations were obtained using standard stars from~\cite{Landolt1992}, 
including SA113-163, SA113-337, SA113-265 and SA92-412.
The full width at half-maximum (FWHM) measured on 2003 EH1 varied from $\sim$0.8\arcsec~to 1.5\arcsec. The sky was photometric on the nights of UT 2013 August 9, 12 and October 2. Data obtained under slightly non-photometric conditions on August 8 were photometrically calibrated using field stars observed on a photometric night.  
An observational log is given in Table~\ref{logEH1}.

%------------------------------------------------------------------------------------------------------------------------

\section{Results}
\label{Obs}

\begin{figure*}[htbp]
\epsscale{1} \plotone{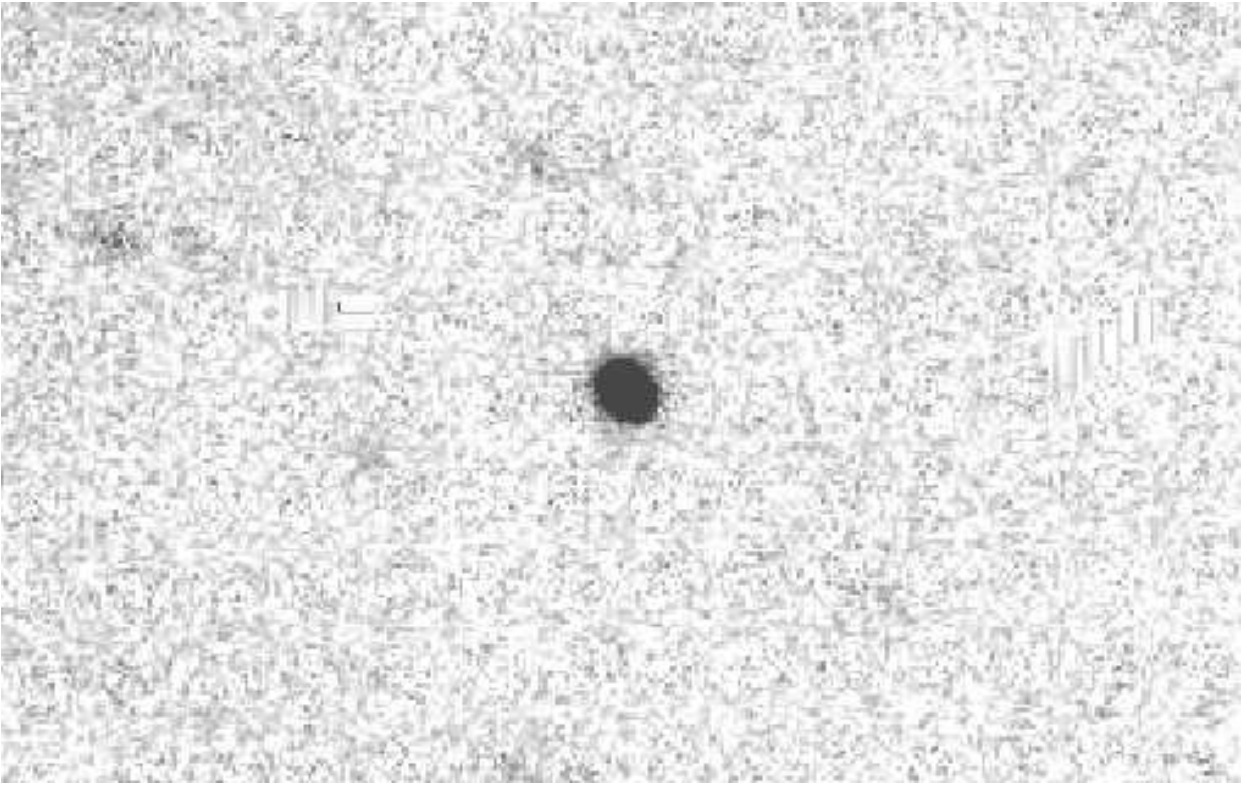} \caption{
The $\it R$-band image of 2003 EH1 
taken by Keck-I 10\,m on UT 2013 October 2.  
The image has total integration time of 360\,s.   
The frame size is 40$^{''}$ $\times$ 25$^{''}$.  
No coma or tail are visible on the object, which has a FWHM of $0.86^{''}$.}
\label{image}
\end{figure*}

Object 2003 EH1 appeared point-like in all image data (see Figure \ref{image}).  
Photometry was performed using synthetic circular apertures projected onto the sky.  
The photometric aperture radius used was twice the FWHM in the image ($\sim$1.6\arcsec~to 3.0\arcsec) 
and the sky background was determined within a concentric annulus having projected 
inner and outer radii of 6.6\arcsec~and 13.2\arcsec, respectively.
Photometric results are listed in Tables~\ref{colorEH1} and \ref{lightphot}.

\subsection{Colors}
\label{ColorC}

\begin{figure*}[htbp]
\epsscale{1.} \plotone{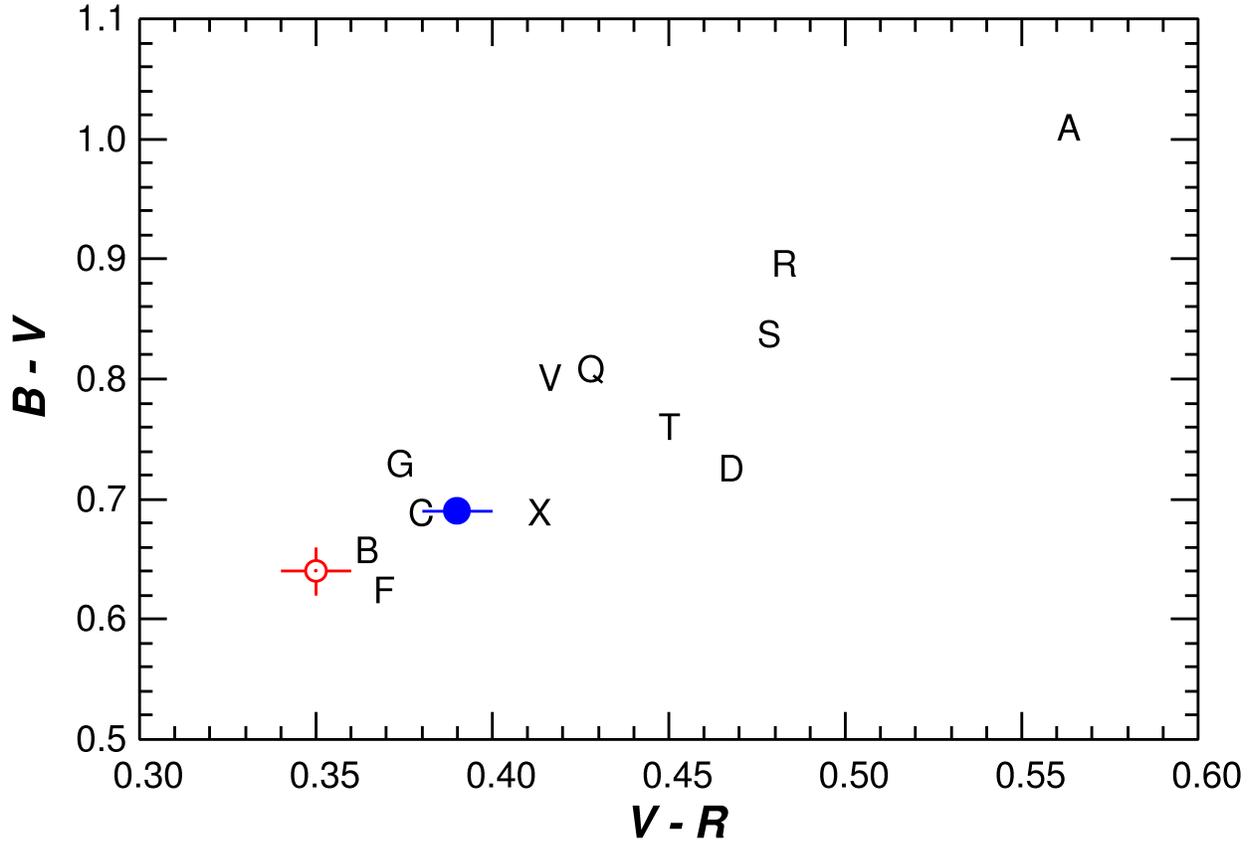} 
\caption{Color plots of ${\it V-R}$ vs. ${\it B-V}$ for 2003 EH1 (blue circle) on weighted mean and 
Tholen taxonomic classifications~\citep{Th84}, as tabulated by~\cite{Da03}.  
The color of the Sun (red circle) is also plotted.  The uncertainty of ${\it B-V}$ for 2003 EH1 is within the circle. }
\label{VRBV}
\end{figure*}

\begin{figure*}[htbp]
\epsscale{1.} \plotone{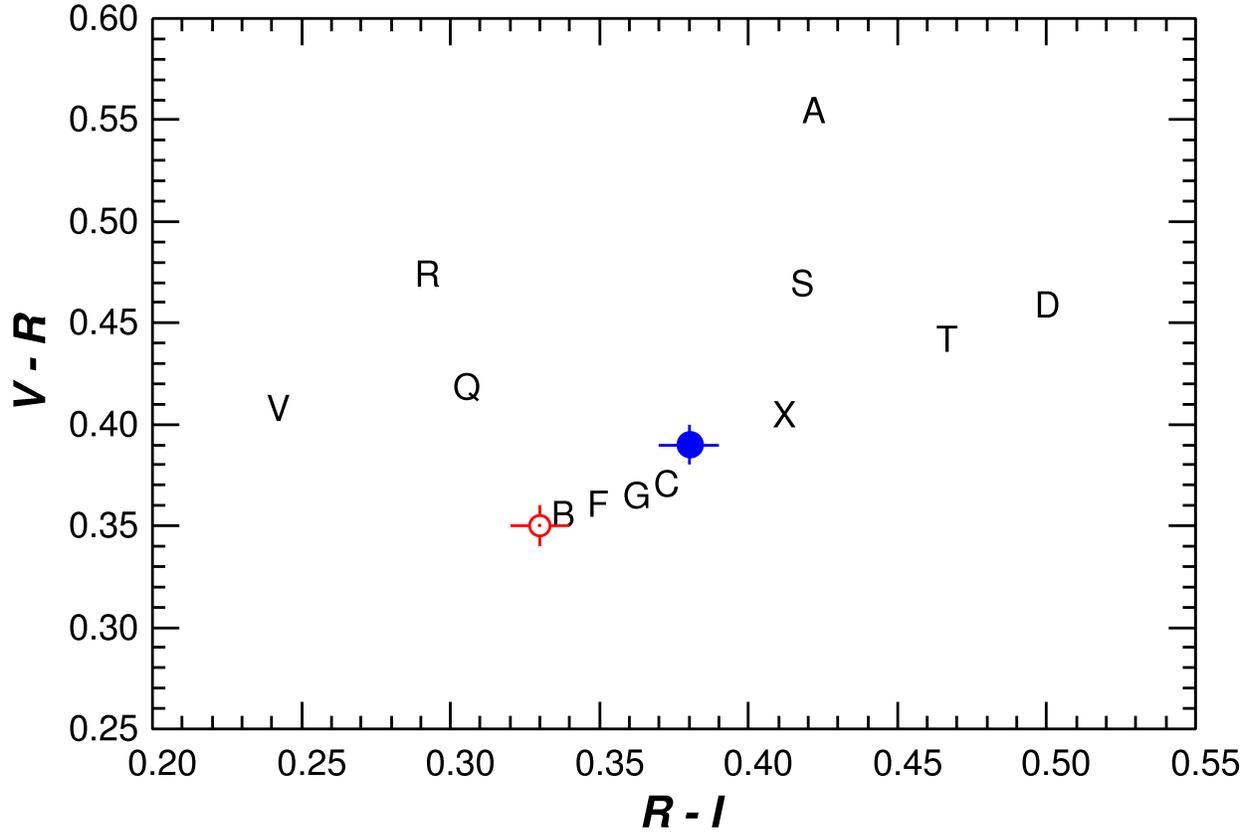} 
\caption{The same as Figure~\ref{VRBV} but in the ${\it R-I}$ vs. ${\it V-R}$ color plane.}
\label{RIVR}
\end{figure*}

The weighted mean colors of 2003 EH1 are ${\it B-V}$ = 0.69$\pm$0.01, ${\it V-R}$ = 0.39$\pm$0.01 
and ${\it R-I}$ = 0.38$\pm$0.01 from N=16 measurements (see Table~\ref{colorEH1}). 
Figures~\ref{VRBV} and~\ref{RIVR} show ${\it V-R}$ vs.~${\it B-V}$ and 
${\it R-I}$ vs.~${\it V-R}$ respectively, together with the Tholen taxonomy classes~\citep{Th84} from \cite{Da03}.  
The V-R data of 2003 EH1 together with the various small body populations and the solar color  
are summarized in Table~\ref{distcolor}. We also list the normalized reflectivity slope, ${\it S'}$ 
$[ \% (1000\,$\rm \AA$)^{-1}]$, measured in the V-R region~\citep{LuuJewitt1990}.

The optical colors of 2003 EH1 are similar to, but slightly redder than,
those of the Sun (Table~\ref{colorEH1}), being most taxonomically compatible 
with those of C-type asteroids (Figs.~\ref{VRBV} and~\ref{RIVR}).     
The V-R  color (0.39$\pm$0.01) is similar to  the weighted mean color of
96P/Machholz (${\it V - R}$ = 0.40$\pm$0.03,  from \cite{Lica00} and \cite{Me04}). 
Table~\ref{distcolor} indicates that 2003 EH1 has a  spectral slope less red than those of 
dead comets, cometary nuclei, Jupiter Trojans and Damocloids, many of which are spectrally 
classified as D-type asteroids~\citep{JewittLuu1990, Fi94, Jewitt2002, Jewitt2004, J05, Fornasier2007, Karlsson2009}.  
On the other hand, 2003 EH1 has  a nearly neutral  spectral slope,  
as do many main belt comets~\citep[MBCs:][]{HJ06} (see Table~\ref{distcolor}).

We note that the colors and ${\it S'}$ of 2003 EH1
are remarkably less red than the average colors of cometary nuclei~\citep{Jewitt2002, Lamy2004}.   
This could be a result of past thermal processing when the object had a perihelion 
far inside Earth's orbit. Indeed, the weighted mean color of 8 near-Sun asteroids having perihelion 
distances $\le$0.25 AU (subsolar temperatures $\ge$800 K) is V-R = 0.36$\pm$0.01 \citep{Jewitt2013}, consistent with the color of EH1.
We conclude that the colors of 2003 EH1 are broadly consistent with  those measured in dead cometary nuclei,  
 presumably as a result of mantling from now-gone activity.

\subsection{Surface Brightness}
\label{SBS}

Here we search for evidence of a coma, which would indicate ongoing mass loss from 2003 EH1.  
We compared the measured surface brightness profile 
with the profiles of a field star nearby and a seeing-convolution model.
Since the non-sidereal motion of 2003 EH1 makes the images of background stars appear 
trailed in the data, the one-dimensional surface brightness profiles were examined using the procedures of \cite{Luu1992}.
To determine the profile, 
we used two $\it R$-band images taken using the Keck-I telescope on UT 2013 October 2 (Table\,\ref{logEH1}), 
without any background contamination.        
The Keck signal-to-noise ratio: S/N $\geq$ 70 - 140 is greater than 
that of the KPNO\,2.1 (S/N $\simeq$ 20 - 30).  
Each image was rotated to bring  the direction of the projected motion of 2003 EH1 to the horizontal, shifted to align the images using fifth-order polynomial interpolation, then combined into a single image 
(total integration time of 360\,sec). 
The resulting  image of 2003 EH1 has a FWHM of 0.86\arcsec,
compatible with the seeing in the individual images used to make the composite. 
The seeing was determined from the point spread function (PSF) of a field star measured perpendicular to the direction of trail
and convolved with ``nucleus plus coma"  comet models.  
In the model images, each of 100 $\times$ 100 pixels, the nucleus was represented as a ``point source" 
located at the central pixel embedded in a circularly symmetric coma of varying activity levels.  
The surface brightness is assumed to decrease inversely with distance from the nucleus, as expected for steady-state, isotropic expansion of a coma. 
The principal parameter $\eta$, is equal to the ratio of the cross sections of the coma to that of the nucleus, with $\eta$ = 0 corresponding to a bare nucleus and $\eta$ = 1 to nucleus and coma having the same cross sections within the projected photometry aperture \citep{Luu1992}.  
The flux density of each pixel in the coma is given by ${\it K}$/${\it r}$, 
where ${\it K}$ is a constant of proportionality and ${\it r}$ is the distance 
from the nucleus in the plane of the sky.  

\begin{figure*}[htbp]
\epsscale{1} \plotone{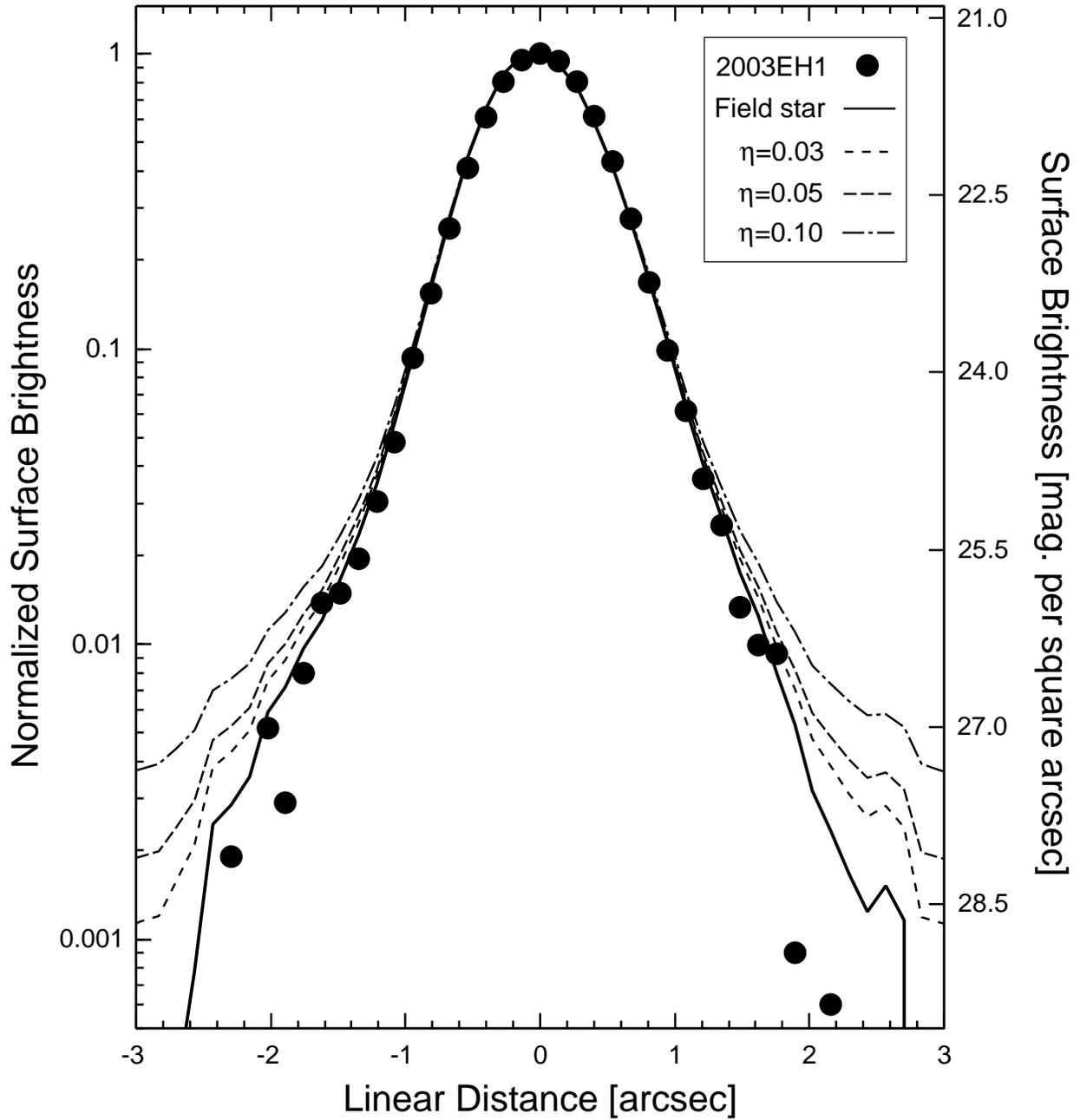} \caption{Normalized $\it R$-band surface brightness profiles of 
2003 EH1, the field star, and seeing-convolution models having coma levels of $\eta$ =0.03, 0.05 and 0.10.  
One unit of the surface brightness of the the asteroid is $\Sigma$=21.3\,mag arcsec$^{-2}$. } 
\label{2D}
\end{figure*}

Figure~\ref{2D} shows surface brightness profiles of 2003 EH1, 
the field star (solid line) and seeing-convolution models 
with coma levels of $\eta$ =0.03, 0.05, 0.10 (dotted lines).  
All profiles are normalized to be unity at the center for comparison.  
The surface brightness profiles of 2003 EH1 and a field star were measured 
in the direction perpendicular to the motion of the asteroid.  
The individual profile, after the sky background subtraction, was averaged 
along the rows over the width of the asteroid and the field star.   
The normalized profiles of the asteroid and the field star are indistinguishable.   
From the figure we set an upper limit on the coma level $\eta$ $\lesssim$ 0.025 $\pm$ 0.007.

A limit to near-nucleus coma can also be set on the basis of simple aperture photometry \citep{Jewitt1984}.
Observations set a limit to the surface brightness, 
$\Sigma$$(\phi)$ mag\,${\rm arcsec^{-2}}$ 
at angular distance $\phi \arcsec$ from the image center.    
If the coma is in steady-state production (i.e.~the surface brightness varies with the inverse of the distance from the nucleus), then 
$m_c$($\phi)$,  the total magnitude of the coma inside radius  $\phi$,  is given by \cite{Jewitt1984} as
\begin{equation}
m_c(\phi) = \Sigma (\phi) -2.5 {\rm log} (2\pi \phi^2).
\label{Jewitt1984}
\end{equation}
From Figure~\ref{2D}, we can be confident that an upper limit to the coma surface brightness at $\phi$=3$\arcsec$ is $\Sigma$$(3'')$ $\sim$ 27\,mag\,${\rm arcsec^{-2}}$.   
Substitution into Equation (\ref{Jewitt1984}) gives $m_{\rm c}$($3.0''$) = 22.6\,magnitude, which  is 2.7\,mag (factor of $\sim$ 12) fainter than the total magnitude 19.9\,mag in the $R$-band.
Therefore, we conclude that the magnitude of coma within a 3$\arcsec$ radius circle is $\leq$ 0.08 of the measured brightness.  This is consistent with, but less stringent than, the limit deduced from the profile-fitting model.

% --------------------------------------------------------------------------------------

\subsection{Size and Active Fractional Area}
\label{sizeS}

To derive the size of 2003 EH1, we used results of the 
${\it R}$-band photometry taken on the nights of UT 2013 August 9 and 12 from KPNO\,2.1 (Table\,\ref{colorEH1})  
and those taken on UT 2013 October 2 from Keck\,10 (${\it R}$ = 20.21$\pm$0.01\,mag and 20.26$\pm$0.02\,mag).
The apparent red magnitude $m_{\rm R}$ was corrected to the absolute red magnitude, $m_{\rm R}(1,1,0)$ using 
\begin{equation}
m_R (1,1,0) = m_R - 5\,{\rm log}(R\, \Delta) -\beta \alpha, 
\label{R}
\end{equation}
where ${\it R}$ and $\Delta$ are the heliocentric and geocentric distances (both in AU), $\alpha$(deg), is the phase angle 
(Observer-asteroid-sun), and $\beta$ is the linear phase coefficient (mag deg$^{-1}$).  
We took $\beta$ = 0.04 mag\,deg$^{-1}$, which is compatible with values measured fior JFC nuclei \citep{Lamy2004}.
We used absolute red magnitude, $m_{\rm R}(1,1,0)$, to calculate the effective object radius in meters, $r_e$, 
using~\cite{Russell1916}
\begin{equation}
r_{\rm e}   = \frac{1.496 \times 10^{8}}{\sqrt{p_{R}}}10^{0.2(R_\odot - m_R (1,1,0))}, 
\label{re} 
\end{equation}
where $R_\odot$ = --27.1 is the apparent red magnitude of the Sun \citep{Cox2000}.  
We adopt the typical value of geometric albedo, $p_{v}(\approx p_{R})$ = 0.04, from   
the visible and thermal (mid-infrared) measurements for JFC nuclei \citep{Lamy2004, Fe13}.  
For the averaged absolute red magnitude $m_{\rm R}(1,1,0)$=15.82$\pm$0.17\,mag, 
Equation\,(\ref{re}) gives $r_e$ = 1950$\pm$150 m, which we approximate as $r_e$ = 2.0$\pm$0.2 km.    
The nucleus, represented by a sphere of this radius and assumed bulk density 
$\rho$ = 2000 kg m$^{-3}$ (the density of the Quadrantid meteoroids \citep{BaK09}), is $M_n \sim$ 6$\times$10$^{13}$ kg.  
This is comparable to, but slightly larger than, the estimated stream mass of (1 to 2)$\times$10$^{13}$ kg.

The asteroid 2003 EH1 shows point-like surface brightness.  
Here we estimate the maximum allowable coma activity.    
Assuming that the water ice still exists and occupies the object surface, 
we estimate limits both to ongoing mass-loss rate and fractional active area on the surface.     
The approximate rate of the isotropic dust ejection from the object is expressed as a function of the 
parameter $\eta$  \citep{Luu1992}: 
\begin{equation}
\frac{dM}{dt} = \frac{1.0 \times 10^{-3} \pi \rho \bar{a} \eta r_{\rm e}^2}{\theta R^{1/2}\Delta}
 \label{massloss}
\end{equation}
where $\rho$=2000 ${\rm kg\,m^{-3}}$ is the assumed bulk density determined by the Quadrantid meteoroids~\citep{BaK09}, 
$\bar{a}$=0.5$\times$$10^{-6}$\,m is the assumed mean grain radius, 
$r_{\rm e}$=1950$\pm$150\,m is the effective radius of 2003 EH1, 
$\theta$ is the reference photometry aperture radius of 30 pixels $(4.05'')$, 
and ${\it R}$ = 2.139 \,AU, ${\it \Delta}$=2.038\,AU given in Table\,\ref{logEH1}.
The estimated limit to the mass loss rate is 
$dM / dt$ $\lesssim$ 2.5$\times$10$^{-2}$ $\,{\rm kg\,s^{-1}}$ 
with $\eta$\,$\lesssim$\,0.025$\pm$0.007.
The $dM / dt$ is converted into the fraction of active area 
on the nucleus surface, $f_A$, using \cite{Luu1992}:
\begin{equation}
f_A = \frac{dM / dt}{4 \pi r^2_{e}\, \mu \,dm/dt},
\label{fraction}
\end{equation}
where $dm/dt$ is the specific sublimation mass loss rate of water in ${\rm kg\,m^{-2}\,s^{-1}}$ 
and $\mu$ = 1 is the assumed dust-to-gas mass ratio \citep{Greenberg1998,Luu1992}. 
(A value $\mu$ = 4$\pm$2 was measured in a recent encounter with JFC 67P/Churyumov-Gerasimenko~\citep[][]{R15}).
The $dm/dt$ is calculated from the energy-balance equation
\begin{equation}
\frac{S_\odot (1-A)}{R^2} = \chi [\epsilon \sigma T^4 + L(T) dm/dt], 
\label{ebala}
\end{equation}
where $S_\odot$ = 1365 W ${\rm m^{-2}}$ is the solar constant, 
${\it R}$ (in AU) is the heliocentric distance, $\epsilon$ = 0.9 is the wavelength-averaged emissivity, 
$\sigma$= 5.67 $\times$ $10^{-8}$ W ${m^{-2}}$ ${K^{-4}}$ 
is the Stephan-Boltzmann constant and $T$ K is the equilibrium temperature.  Quantity
A is the Bond albedo, defined by $A$ = $p_v\,q$ = 0.012, where $p_v$=0.04 \citep{Lamy2004, Fe13}
and $q \sim$ 0.3 is the phase integral 
determined from cometary nuclei and Jupiter Trojan asteroids \citep{Fernandez2003, Buratti2004}.
The latent heat of sublimation for water at temperature $T$ (in K) is given by 
$L(T)$ = (2.875 $\times$ $10^6$) -- (1.111 $\times$ $10^{3}$)$T$ in J ${\rm kg^{-1}}$, 
taking the  polynomial fit to the thermodynamic data in \cite{Delsemme1971}.

The dimensionless parameter $\chi$ represents the ratio of 
the effective cross-section for emission of thermal radiation from the nucleus to that for absorption of solar power.  
The lowest value, $\chi$=1, corresponds to subsolar ice on a non-rotating object, 
while the highest value, $\chi$=4, corresponds to  an isothermal, spherical nucleus.  
For comet-like objects, the night-side thermal radiation is negligible (i.e. day-side emission only)
due to the low thermal diffusivity of the surface layers, suggesting the intermediate value, $\chi$=2, is appropriate for 
providing a maximum active fractional area and minimum specific mass ross rate \citep{Fe13, LJ15}.  
However, since we are interested in obtaining a limit to $f_A$, we assume the lowest possible surface temperatures (corresponding to the isothermal case, $\chi$ = 4) and find $dm/dt$ = 7.5$\times$10$^{-6}$ kg m$^{-2}$ s$^{-1}$ and $T$ = 180 K at $R$ = 2.139 AU, by Equation (\ref{ebala}).  To supply 2.5$\times$10$^{-2}$ kg s$^{-1}$ would require an exposed patch of ice on the surface having area 3300 m$^2$, corresponding to $f_A \lesssim$ 10$^{-4}$ by Equation (\ref{fraction}).  
This fraction 
is  smaller by an order of magnitude than is characteristic of even low activity JFC nuclei \citep{AHearn1995}.

% kai=4, $f_A \sim$ 7.0$\pm$1.9 $\times$ $10$^{-5}

% --------------------------------------------------------------------------------------

\subsection{Rotational Period and Shape}
\label{ps}

\begin{figure*}[htbp]
\epsscale{1.} \plotone{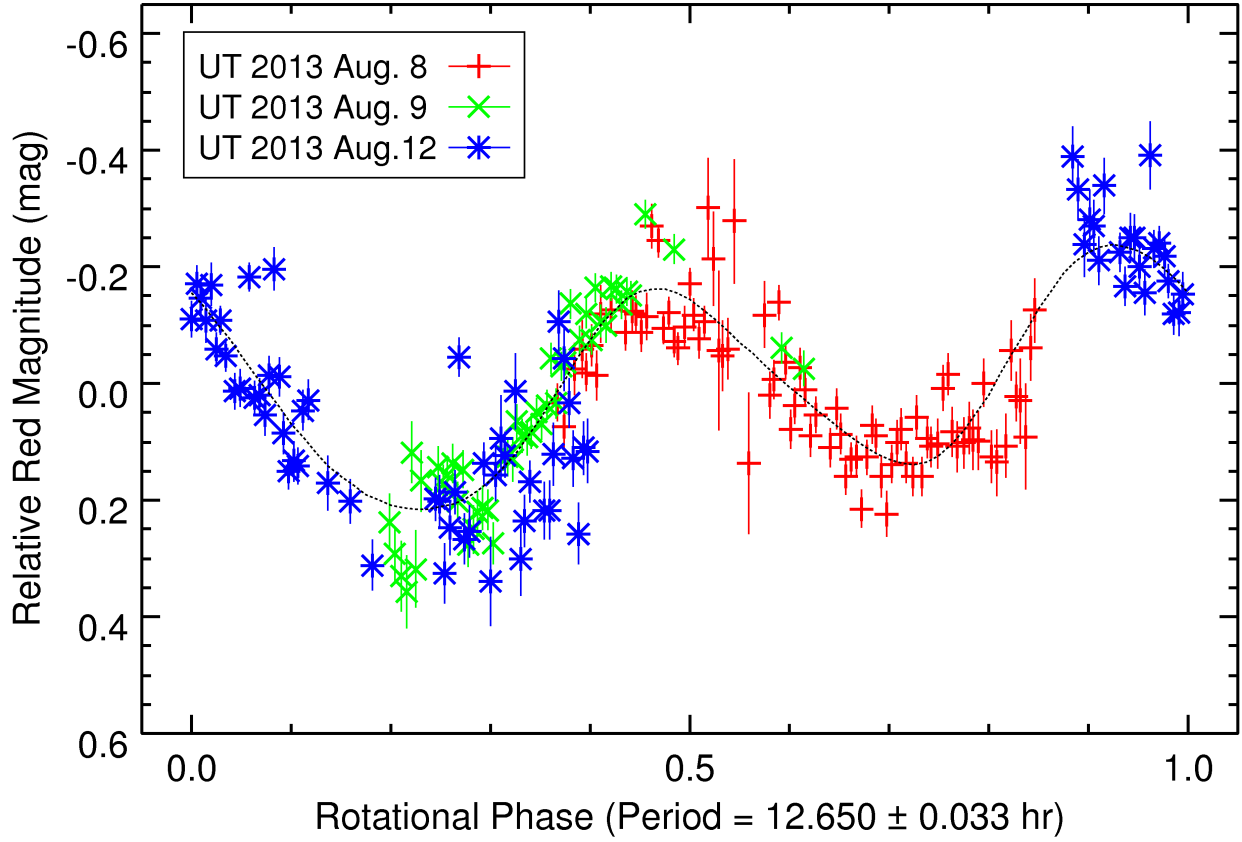} \caption{${\it R}$-band photometry of 2003 EH1 observed on UT 2013 August 8, 9 and 12
, phased to the two-peaked period $P_{\rm rot}$=12.650$\pm$0.033\,hr.  
Dotted curve displays fitting result having the maximum photometric range $\Delta m_{\rm R}$= 0.44 $\pm$ 0.01\,mag.}
\label{LC}
\end{figure*}

To search for the rotation period for 2003 EH1, 
we used a spectral analysis technique that employs the 
Discrete Fourier Transform (DFT) algorithm \citep[][]{Lomb1976,Scargle1982}
on the relative ${\it R}$-band time-series photometric data (Table~\ref{lightphot}).
The DFT analysis evaluates the spectral power as a function of angular frequency 
using the fitting quality at a given frequency in the data. 
The maximum power at the frequency indicates the highest significance level, 
reflecting the most convincing solution for the periodicity. 
The light curve shape is presumed to be two-peaked as seen in 
most small bodies in the Solar System, implying 
 elongated body shape.
The fitting solution for the two-peaked rotational period is $P_{\rm rot}$=12.650\,hr. 
The uncertainty on the period is computed using the equation given by \cite{GF85} 
\begin{equation}
\frac{\Delta f}{f} = \left[ \frac{0.0256}{(fT)^4} + \frac{0.5625 \sigma^2}{n (fT)^2 A^2}  \right]^{1/2}, 
\label{err}
\end{equation}
where $\Delta$$f$ is the root-mean square error, $f$ is the number of cycles per day (24\,hrs), 
$T$ is the observing period (in days), $A$ is the signal amplitude,  $n$ is the number of measurements 
and $\sigma^2$ is the variance of the data. 
Substituting $f$ = 1.8972 (= 24\,hr/$P_{\rm rot}$), $T$ = 4.2299, $A$=0.44\,mag, $n$ = 205 and $\sigma^2$=0.0025, 
we obtain $\Delta f/f$ $\sim$ 0.26\%, namely, the uncertainty on the period is $\pm$0.033\,hr. 
The phased light curve with this period, $P_{\rm rot}$=12.650$\pm$0.033\,hr, is shown in Figure~\ref{LC}.

The fitted model for the light curve finds the maximum photometric range 
of 2003 EH1 is $\Delta m_{\rm R}$= 0.44 $\pm$ 0.01, 
which gives a lower limit to the intrinsic axis ratio, ${\it a/b}$, 
between long axis ${\it a}$ and short axis ${\it b}$.
Assuming the object's rotational axis is perpendicular to our line of sight, 
the ratio is expressed as $a/b = 10^{0.4 \Delta m_{\rm R}}$.  
We find ${\it a/b}$ = 1.50 $\pm$ 0.01.  In practice, this is a lower limit to $a/b$ because the rotation axis may not be perpendicular to the line of sight.
Our observations of 2003 EH1 are consistent with the shapes of typical cometary nuclei, which tend to be elongated (${\it a/b}$$\ge$ 1.5 \citep{Jewitt2004}) relative to asteroids of comparable size.  The slow rotation and modest $a/b$ do not present any threat to the rotational stability of 2003 EH1 for bulk densities $>$100 kg m$^{-3}$, even assuming zero tensile strength.

Non-central outgassing (mass loss) can generate torques that 
change the angular momentum of the nucleus and 
which can drive an object into an excited rotational energy state.
We estimated  
the timescale for rotational excitation of 2003 EH1 assuming continuous mass loss at the maximum rate allowed by our data and using the formalism described in \cite{Jewitt1997}.  With values of the dimensionless moment arm for the torque in the range 10$^{-3}$ to 10$^{-1}$, we obtain excitation timescales in the range from 10$^5$ to 10$^7$ yr.  These are long compared to the few$\times$10$^4$ yr active lifetimes of JFC comets \citep{LevisonDuncan1997}, suggesting that rotational excitation of 2003 EH1 is unlikely, at least given the present activity state.

\subsection{Mantle Formation}

Rubble mantles in comets consist of refractory blocks that are large enough 
not to be ejected by outgassing drag forces  against the gravity of the nucleus, although cohesion also likely plays a role.  The timescale for growth of a cohesionless rubble mantle in the presence of a sublimating ice surface is given by Jewitt (2002).  From Figure 5 of that paper, we read that the mantling time for a 2 km nucleus between 1 and 5 AU from the Sun is in the range 0.3 $\lesssim \tau \lesssim$ 100 yr.  Even the upper limit to the timescale is short compared to the timescale of the dynamical evolution of 2003 EH1, showing that mantle formation is likely and explaining the very low (or absent) present-day mass loss.  Given that 2003 EH1 has followed a complicated and rapidly changing dynamical path, including recent close-passages by the Sun, it is likely that the existing rubble mantle reflects depletion of near-surface volatiles occurring at higher temperatures than those that now prevail.  

The timescale for heat to conduct across the radius of the nucleus, $r_e$, 
is of order $\tau_h \sim r_e^2  / \kappa$. 
With $r_e$ = 2 km and thermal diffusivity $\kappa$ = 10$^{-8}$ to 10$^{-7}$ m$^2$ s$^{-1}$ (as appropriate for a porous dielectric material), we find $\tau_h \sim$ 10$^6$ to 10$^7$yr.
The $\tau_h$ exceeds the dynamical lifetime of JFC comets $\tau_{\rm JFC}$ $\sim$ 10$^5$yr \citep[][]{LevisonDuncan1994} by one or more orders of magnitude, showing that the heat from the Sun would not reach deep interior of the asteroid during the time spent in the inner solar system.
Therefore, we conclude that it is very plausible that 2003 EH1 retains volatiles in its deep interior, but that it is inactive during most of its orbit owing to 
the recent (and probably recurring) formation of a rubble mantle.

%----------------------------------------------------------------------------------------
\section{Discussion}
\label{discuss}

% Stream Mass
As noted earlier, the Quadrantid core stream is estimated from
dynamical spreading to be 200 to 500 years in age \citep{Jennis04, WB05, Abedin15}.  Steady mass loss at the maximum rates allowed by the optical data,
namely 2.5 $\times$ $10^{-2}$ kg ${\rm s^{-1}}$, would deliver only
about (1.6 - 3.9)$\times$ $10^8$ kg in 200 - 500 yr, even if these rates were sustained all around the orbit (which itself seems unlikely). 
For comparison, the total mass of the meteoroids in the Quadrantid core stream  
is estimated to be about $10^{13}$\,kg~\citep{jennis06}, which has been 
updated from earlier estimates of $\leq$ $10^{11-12}$\,kg \citep{HM89, jenn94,Jennies97}. 
We conclude that the current  production rates from 2003 EH1 are about five orders of magnitude too small to 
supply the mass of the core Quadrantid stream.  This result is perhaps not surprising, given the current mis-match between the orbits of 2003 EH1 and the Quadrantid stream  \citep{WB05}.

Could the core stream meteoroids have been released from 2003 EH1 a few centuries ago, when the perihelion was substantially smaller?  For example, 200 to 500 years ago, the perihelion distance was $\sim$0.7 to 0.9 AU \citep{Jennis04,WB05}. We solved Equation~(\ref{ebala}) to 
find hemispherically averaged specific mass loss rates (2.8 - 4.9) $\times$10$^{-4}$ kg m$^{-2}$ s$^{-1}$ at these distances, only 2 to 3 times larger than at 1.2 AU.  
Thus, perihelion variations alone are not sufficient to account for the mass of the Quadrantids.  Within the context of the equilibrium sublimation model, only by changing the active fraction, $f$, can the production rates and the stream mass be reconciled.  For example, setting $dm/dt$ = 4.9$\times$10$^{-4}$ kg m$^{-2}$ s$^{-1}$ and $f_A$ = 1 in Equation (\ref{fraction}) we find that the stream mass could be supplied by equilibrium sublimation in $\sim$30 years.  We consider it more likely that the injection of mass to the meteoroid stream occurred out of equilibrium, perhaps by a volatile-driven process related to cometary outbursts or break-ups, and triggered by deep penetration of conducted heat into the ice-rich interior of this body.

% 1yr = 3.1556 x 10^7 second

%% Sun-grazing phase
Intense solar heating can cause fracturing and dust production through thermal fracture and desiccation.
For example, asteroid (3200) Phaethon, the parent body of the 
Geminid meteoroid stream,  has shown  recurrent 
activity around its perihelion ${\it q}$$\sim$0.14\,AU \citep{JL10,LJ13,Jetal13} 
where the surface temperature reaches 750\,K $\le$ $\it T$ $\le$ 1100\,K~\citep{Oh09}.
Phaethon is essentially a ``rock comet" and the activity is caused by  
the production of small dust particles with radii $\sim$ 1\,$\micron$ 
due to thermal fracture and decomposition cracking of hydrated minerals (not sublimation of ice). 
Since  2003 EH1 recently possessed similarly small perihelia~\citep[][]{Nes13b,FJ14}, thermal fracture and surface desiccation may likewise be expected.
At its smallest perihelion, $\it q$ $\sim$ 0.12\,AU, 
we estimate surface temperatures  
800\,K $\le$ $\it T$ $\le$ 1200\,K on 2003 EH1.   However, as on (3200) Phaethon, the particles produced this way should be of micron size and swept from the nucleus by solar radiation pressure \citep{Jetal13,J15IV}, so that they do not contribute to the meteoroid streams of either body.

%% Na depletion from meteor 
Spectroscopic measurements of the Na contents in the meteoroid streams 
are also suggestive of thermal processing of the parent bodies.
The Geminid meteoroids show extreme diversity in their Na abundance, from
strong depletion to near sun-like Na content \citep{Hv73,kasuga2005,Jiri2005}.  
Presumably, this compositional diversity reflects different  thermal modification on  Phaethon 
(or perhaps the larger sized precursor body) itself
~\citep{kasuga2006,Ohtsuka2006,jewitt2006,Ohtsuka2008,KJ08,K09,Oh09, CB09}.   
For the Quadrantid meteoroids, the measured line intensity ratios show that 
Na is less depleted than in  the 
majority of Geminid meteoroids~\citep[][]{Koten06,Bo10}.  
This may imply  less thermal modification on 2003 EH1 even though it recently had perihelion distances smaller than Phaethon's.  
Alternatively, the Quadrantid meteoroids could be released from sub-surface regions on 2003 EH1 deeper than a thermal skin depth and thereby have escaped the most severe thermal effects~\citep[][]{Koten06}.

%----------------------------------------------------------------------------------------

\section{Summary}

Optical observations of suggested  Quadrantid  stream parent 2003 EH1 lead to the following results.

\begin{enumerate}
\item The absolute red magnitude, ${\it m}_{\rm R}$(1,1,0)=15.82$\pm$0.17\,mag., corresponds to  
         effective radius  $r_{\rm e}$=2.0$\pm$0.2\,km assuming  red geometric albedo $p_{\rm R}$=0.04.  
         The ratio of the nucleus mass to the Quadrantid stream mass is $\sim$3 to 6, 
         although uncertainty remains because both masses are approximate.  

\item The surface brightness profile is point-like, limiting 
         the fractional light scattered by steady-state, near-nucleus coma to $\leq$ 2.5 \%. The maximum mass loss rate deduced from a model fitted to the profile is 
         $\sim$ 2.5$\times$10$^{-2}$ kg s$^{-1}$.  Water ice can occupy a fraction of the surface no larger than $f_A <$ 10$^{-4}$.  
         
\item  The two-peaked rotational  light curve has period $P_{\rm rot}$=12.650$\pm$0.033\,hr. The photometric range,  $\Delta m_{\rm R}$= 0.44$\pm$0.01 ,
         indicates a minimum axis ratio of 1.50 $\pm$ 0.01. 
       
 \item  The optical colors (${\it B-V}$ = 0.69$\pm$0.01, ${\it V-R}$ = 0.39$\pm$0.01, 
          and ${\it R-I}$ = 0.38$\pm$0.01) are slightly redder than the Sun and consistent with the mean colors of dead or dormant cometary nuclei.

\item  Current dust production from 2003 EH1 is orders of magnitude  too small to supply the mass of the Quadrantid core meteoroid stream in the  200-500 year dynamical lifetime.  If 2003 EH1 is the source of the Quadrantids, we infer that mass must be delivered episodically, not in steady-state.
\end{enumerate}

\acknowledgments
 We acknowledge Lusine Kamikyan for assistance with the observations at KPNO\,2.1 telescope.   
 TK is grateful to support for this work provided by Tomoko Arai and Takafumi Matsui, collaborated with 
 the International Space Station METEOR project at Planetary Exploration Research Center, Chiba Institute of Technology. 
 TK also thanks financial supports from National Astronomical Observatory of Japan and 
 a Young Researcher Overseas Visit Program (2013), The Graduate University for Advanced Studies.        
 DJ appreciates support of this work from NASA's Solar System observations program.
 We thank Dianne Harmer, Beatrice Mueller, Anna Daniel
 and other members in the KPNO\,2.1 telescope team for their 
 help in planning and scheduling the observations.  We thank Joel Aycock for operating the Keck telescope and the anonymous referee for comments.  Some of the data presented herein were obtained at the W.M. Keck Observatory, which is operated as a scientific partnership among the California Institute of Technology, the University of California and the National Aeronautics and Space Administration. The Observatory was made possible by the generous financial support of the W.M. Keck Foundation.

%%%%%%%%%%%%%%%%%%%%%%%%%%%%%%%%%%%%%%%%%%%%%%
%% Tables 

\clearpage
\begin{deluxetable}{lccccccccc}
\tabletypesize{\scriptsize}
\tablecaption{Observation Log \label{logEH1}}
\tablewidth{0pt}
\tablehead{
     UT Date & Telescope\tablenotemark{a}  & Integration & Filter\tablenotemark{b} &  $R$\tablenotemark{c}  & $\Delta$\tablenotemark{d} & $\alpha$\tablenotemark{e} \\
             &                             &  (second)        &        &         (AU)           &           (AU)            &   (deg) }
\startdata
2013 Aug. 8  & KPNO\,2.1  & 180    &  83\,$R$                              & 2.5427 & 2.1059 & 22.81\\
2013 Aug. 9  & KPNO\,2.1  & 180    &  1\,$B$, 48\,$R$                      & 2.5357 & 2.1033 & 22.91\\
             &            & 200    &  2\,$B$                               &        &        &      \\
             &            & 120    &  1\,$V$                               &        &        &      \\
             &            & 140    &  3\,$V$                               &        &        &      \\
2013 Aug. 12 & KPNO\,2.1  & 200    &  3\,$B$                               & 2.5145 & 2.0964 & 23.19\\
             &            & 140    &  3\,$V$                               &        &        &      \\
             &            & 180    &  74\,$R$                              &        &        &      \\
             &            & 300    &  3\,$I$                               &        &        &      \\
2013 Oct. 2  & Keck\,10   & 260    &  $R$                                  & 2.1390 & 2.0383 & 27.59\\ 
             &            & 100    &  $R$                                  &        &        &      \\ 
\enddata
%% Text for table notes should follow after the \enddata but before
%% the \end{deluxetable}. Make sure there is at least one \tablenotemark
%% in the table for each \tablenotetext.
\tablenotetext{a}{KPNO\,2.1 = Kitt Peak 2.1\,m telescope, Keck\,10 = 10\,m Keck-I telescope.}
\tablenotetext{b}{Filter and number of images.} 
\tablenotetext{c}{Heliocentric distance.} 
\tablenotetext{d}{Geocentric distance.} 
\tablenotetext{e}{Phase angle.}
\end{deluxetable}
 % table 1
% label logEH1
\clearpage

\begin{deluxetable}{ccccccc}
\tabletypesize{\scriptsize}
\tablewidth{0pt}
\tablecaption{Color photometry (UT 2013 August 9 and 12) \label{colorEH1}}
\tablehead{
\colhead{N}   & \colhead{Midtime} & \colhead{$B - R$} & \colhead{$V - R$} &\colhead{$R - I$} & \colhead{$B - V$$^{a}$}  & \colhead{$R$}}
\startdata
% Aperture diameter fwhm x 4 with R-interpolation					        			
%    midtime  	     B-R          	    V-R	                  R-I               B-V             R(average)	 	
1  & 30.87809 &			&			&		      &                & 20.21$\pm$0.02  \\ 
2  & 30.94424 &	1.10$\pm$0.01	&			&		      &                & 20.17$^{b}$	 \\  
3  & 31.00062 &			&	0.34$\pm$0.02	&		      & 0.76$\pm$0.02  & 20.16$^{b}$	 \\  
4  & 31.05442 &			&			&		      &                & 20.09$\pm$0.02  \\ 
5  & 31.30880 &	1.02$\pm$0.01	&			&		      &                & 20.13$^{b}$	 \\  
6  & 31.36776 &			&	0.45$\pm$0.02	&		      & 0.57$\pm$0.02  & 20.13$^{b}$	 \\  
7  & 31.42471 &			&			&		      &                & 20.16$\pm$0.02  \\ 
8  & 32.73436 &			&	0.31$\pm$0.02	&		      & 0.71$\pm$0.02  & 20.30$^{b}$	 \\  
9  & 32.79421 &			&			&		      &                & 20.30$\pm$0.02  \\ 
10 & 32.95557 &	1.06$\pm$0.01	&			&		      &                & 20.34$^{b}$	 \\  
11 & 33.01491 &			&	0.48$\pm$0.02	&		      & 0.58$\pm$0.02  & 20.35$^{b}$	 \\  
12 & 33.07137 &			&			&		      &                & 20.36$\pm$0.02  \\ 
13 & 102.67640&			&			&		      &                & 20.33$\pm$0.02  \\ 
14 & 102.80208&	1.11$\pm$0.02	&			&		      &                & 20.40$^{b}$	 \\  
15 & 102.86185&			&	0.39$\pm$0.02	&		      & 0.72$\pm$0.03  & 20.43$^{b}$	 \\  
16 & 102.92412&			&			&		      &                & 20.48$\pm$0.02  \\
17 & 103.00632&			&			&	0.41$\pm$0.04 &                & 20.48$^{b}$	 \\  
18 & 103.09013&	1.19$\pm$0.02	&			&		      &                & 20.51$^{b}$	 \\  
19 & 103.15221&			&	0.42$\pm$0.01	&		      & 0.77$\pm$0.02  & 20.53$^{b}$	 \\  
20 & 103.21338&			&			&		      &                & 20.52$\pm$0.02  \\
21 & 103.29392&			&			&	0.48$\pm$0.03 &                & 20.58$^{b}$	 \\  
22 & 103.37791&	1.02$\pm$0.02	&			&		      &                & 20.60$^{b}$	 \\  
23 & 103.43949&			&	0.27$\pm$0.02	&		      & 0.75$\pm$0.03  & 20.62$^{b}$	 \\  
24 & 103.49812&			&			&		      &                & 20.64$\pm$0.03  \\
25 & 103.57827&			&			&	0.37$\pm$0.01 &                & 20.65$^{b}$	 \\  
Average Colors$^{c}$& &1.07$\pm$0.01   &       0.39$\pm$0.01   &       0.38$\pm$0.01 & 0.69$\pm$0.01 &           \\
Solar Colors$^{d}$  & &0.99$\pm$0.02   &       0.35$\pm$0.01   &       0.33$\pm$0.01 & 0.64$\pm$0.02 &            \\
%
%\cutinhead{This is a cut-in head}
%\sidehead{I'm a side head:}
\enddata
\tablenotetext{a}{Calculated from $B-R$ and $V-R$ in this Table.} 
\tablenotetext{b}{Apparent $R$-band magnitude interpolated from the light curve data.}
\tablenotetext{c}{The weighted mean of measurements.}
\tablenotetext{d}{\cite{H06}}
%\tablenotetext{a}{}
%% You can append references to a table using the \tablerefs command.
%\tablerefs{
%(1) Barbuy, Spite, \& Spite 1985; (2) Bond 1980; (3) Carbon et al. 1987}
\end{deluxetable}

 % table 2
% label colorEH1
\clearpage

\begin{deluxetable}{lccr}
\tabletypesize{\scriptsize}
\tablewidth{0pt}
\tablecaption{${\it R}$-band photometry on KPNO\,2.1 \label{lightphot}}
\tablehead{ \colhead{N} & \colhead{Date (UT 2013)}& \colhead{Midtime\tablenotemark{a}}& \colhead{Relative: ${\it R}$\tablenotemark{b}} }        
\startdata
1  & Aug.8   & 4.65021   &    0.033 $\pm$ 0.033    \\
2  & Aug.8   & 4.72038   &    0.073 $\pm$ 0.038    \\
3  & Aug.8   & 4.86321   &   -0.025 $\pm$ 0.045    \\
4  & Aug.8   & 4.95241   &   -0.059 $\pm$ 0.050    \\
5  & Aug.8   & 5.01383   &   -0.019 $\pm$ 0.032    \\
6  & Aug.8   & 5.07517   &   -0.066 $\pm$ 0.036    \\
7  & Aug.8   & 5.13679   &   -0.014 $\pm$ 0.043    \\
8  & Aug.8   & 5.19821   &   -0.120 $\pm$ 0.049    \\
9  & Aug.8   & 5.32672   &   -0.127 $\pm$ 0.035    \\
10 & Aug.8   & 5.51110   &   -0.087 $\pm$ 0.031    \\
11 & Aug.8   & 5.58944   &   -0.125 $\pm$ 0.036    \\
12 & Aug.8   & 5.65078   &   -0.119 $\pm$ 0.033    \\
13 & Aug.8   & 5.71217   &   -0.087 $\pm$ 0.033    \\
14 & Aug.8   & 5.77379   &   -0.115 $\pm$ 0.040    \\
15 & Aug.8   & 5.83524   &   -0.270 $\pm$ 0.038    \\
16 & Aug.8   & 5.93243   &   -0.245 $\pm$ 0.028    \\
17 & Aug.8   & 5.99379   &   -0.095 $\pm$ 0.027    \\
18 & Aug.8   & 6.05513   &   -0.121 $\pm$ 0.027    \\
19 & Aug.8   & 6.11675   &   -0.073 $\pm$ 0.027    \\
20 & Aug.8   & 6.17812   &   -0.060 $\pm$ 0.028    \\
21 & Aug.8   & 6.25545   &   -0.097 $\pm$ 0.036    \\
22 & Aug.8   & 6.31677   &   -0.172 $\pm$ 0.027    \\
23 & Aug.8   & 6.37812   &   -0.118 $\pm$ 0.028    \\
24 & Aug.8   & 6.43947   &   -0.076 $\pm$ 0.032    \\
25 & Aug.8   & 6.50084   &   -0.105 $\pm$ 0.027    \\
26 & Aug.8   & 6.56360   &   -0.301 $\pm$ 0.085    \\
27 & Aug.8   & 6.62498   &   -0.214 $\pm$ 0.081    \\
28 & Aug.8   & 6.68636   &   -0.057 $\pm$ 0.137    \\
29 & Aug.8   & 6.74799   &   -0.048 $\pm$ 0.061    \\
30 & Aug.8   & 6.80937   &   -0.059 $\pm$ 0.053    \\
31 & Aug.8   & 6.88904   &   -0.278 $\pm$ 0.106    \\
32 & Aug.8   & 7.07318   &    0.137 $\pm$ 0.122    \\
33 & Aug.8   & 7.27336   &   -0.118 $\pm$ 0.058    \\
34 & Aug.8   & 7.33471   &    0.020 $\pm$ 0.041    \\
35 & Aug.8   & 7.39608   &   -0.007 $\pm$ 0.034    \\
36 & Aug.8   & 7.45743   &   -0.139 $\pm$ 0.030    \\
37 & Aug.8   & 7.53773   &   -0.037 $\pm$ 0.030    \\
38 & Aug.8   & 7.59908   &    0.079 $\pm$ 0.034    \\
39 & Aug.8   & 7.66049   &    0.038 $\pm$ 0.032    \\
40 & Aug.8   & 7.72209   &   -0.027 $\pm$ 0.035    \\
41 & Aug.8   & 7.78347   &    0.010 $\pm$ 0.031    \\
42 & Aug.8   & 7.85684   &    0.089 $\pm$ 0.037    \\
43 & Aug.8   & 7.91815   &    0.054 $\pm$ 0.031    \\
44 & Aug.8   & 8.10221   &    0.110 $\pm$ 0.033    \\
45 & Aug.8   & 8.18501   &    0.043 $\pm$ 0.035    \\
46 & Aug.8   & 8.24638   &    0.087 $\pm$ 0.035    \\
47 & Aug.8   & 8.30772   &    0.159 $\pm$ 0.031    \\
48 & Aug.8   & 8.36909   &    0.131 $\pm$ 0.031    \\
49 & Aug.8   & 8.43053   &    0.125 $\pm$ 0.035    \\
50 & Aug.8   & 8.50853   &    0.215 $\pm$ 0.033    \\
51 & Aug.8   & 8.56987   &    0.126 $\pm$ 0.030    \\
52 & Aug.8   & 8.63125   &    0.071 $\pm$ 0.032    \\
53 & Aug.8   & 8.69263   &    0.090 $\pm$ 0.030    \\
54 & Aug.8   & 8.75424   &    0.159 $\pm$ 0.036    \\
55 & Aug.8   & 8.82486   &    0.224 $\pm$ 0.039    \\
56 & Aug.8   & 8.88624   &    0.138 $\pm$ 0.032    \\
57 & Aug.8   & 8.94772   &    0.101 $\pm$ 0.038    \\
58 & Aug.8   & 9.00907   &    0.079 $\pm$ 0.036    \\
59 & Aug.8   & 9.07043   &    0.136 $\pm$ 0.035    \\
60 & Aug.8   & 9.14868   &    0.159 $\pm$ 0.034    \\
61 & Aug.8   & 9.21012   &    0.059 $\pm$ 0.038    \\
62 & Aug.8   & 9.27147   &    0.159 $\pm$ 0.034    \\
63 & Aug.8   & 9.33284   &    0.095 $\pm$ 0.034    \\
64 & Aug.8   & 9.39421   &    0.108 $\pm$ 0.035    \\
65 & Aug.8   & 9.47617   &    0.103 $\pm$ 0.042    \\
66 & Aug.8   & 9.53757   &    0.008 $\pm$ 0.040    \\
67 & Aug.8   & 9.59891   &   -0.016 $\pm$ 0.035    \\
68 & Aug.8   & 9.66027   &    0.083 $\pm$ 0.043    \\
69 & Aug.8   & 9.72165   &    0.108 $\pm$ 0.044    \\
70 & Aug.8   & 9.80528   &    0.102 $\pm$ 0.044    \\
71 & Aug.8   & 9.86663   &    0.076 $\pm$ 0.045    \\
72 & Aug.8   & 9.92799   &    0.096 $\pm$ 0.051    \\
73 & Aug.8   & 9.98936   &    0.098 $\pm$ 0.050    \\
74 & Aug.8   & 10.05075  &   -0.001 $\pm$ 0.042    \\
75 & Aug.8   & 10.15400  &    0.126 $\pm$ 0.045    \\
76 & Aug.8   & 10.21533  &    0.135 $\pm$ 0.057    \\
77 & Aug.8   & 10.33809  &    0.107 $\pm$ 0.055    \\
78 & Aug.8   & 10.39944  &   -0.056 $\pm$ 0.052    \\
79 & Aug.8   & 10.46384  &    0.023 $\pm$ 0.043    \\
80 & Aug.8   & 10.52545  &    0.030 $\pm$ 0.073    \\
81 & Aug.8   & 10.58685  &    0.091 $\pm$ 0.091    \\
82 & Aug.8   & 10.64823  &   -0.062 $\pm$ 0.057    \\
83 & Aug.8   & 10.70965  &   -0.126 $\pm$ 0.055    \\
84 & Aug.9   & 27.81903  &    0.237 $\pm$ 0.049    \\
85 & Aug.9   & 27.88119  &    0.293 $\pm$ 0.049    \\
86 & Aug.9   & 27.96539  &    0.330 $\pm$ 0.060    \\
87 & Aug.9   & 28.02678  &    0.357 $\pm$ 0.063    \\
88 & Aug.9   & 28.08838  &    0.118 $\pm$ 0.052    \\
89 & Aug.9   & 28.14975  &    0.318 $\pm$ 0.067    \\
90 & Aug.9   & 28.21116  &    0.167 $\pm$ 0.053    \\
91 & Aug.9   & 28.42676  &    0.143 $\pm$ 0.036    \\
92 & Aug.9   & 28.48830  &    0.173 $\pm$ 0.033    \\
93 & Aug.9   & 28.54993  &    0.158 $\pm$ 0.034    \\
94 & Aug.9   & 28.61124  &    0.137 $\pm$ 0.034    \\
95 & Aug.9   & 28.67285  &    0.201 $\pm$ 0.033    \\
96 & Aug.9   & 28.75240  &    0.147 $\pm$ 0.033    \\
97 & Aug.9   & 28.81376  &    0.277 $\pm$ 0.037    \\
98 & Aug.9   & 28.87517  &    0.250 $\pm$ 0.041    \\
99 & Aug.9   & 28.93656  &    0.221 $\pm$ 0.038    \\
100& Aug.9   & 28.99796  &    0.213 $\pm$ 0.038    \\
101& Aug.9   & 29.06066  &    0.217 $\pm$ 0.035    \\
102& Aug.9   & 29.12197  &    0.275 $\pm$ 0.036    \\
103& Aug.9   & 29.30653  &    0.126 $\pm$ 0.034    \\
104& Aug.9   & 29.37069  &    0.131 $\pm$ 0.038    \\
105& Aug.9   & 29.43208  &    0.066 $\pm$ 0.030    \\
106& Aug.9   & 29.49344  &    0.090 $\pm$ 0.029    \\
107& Aug.9   & 29.55509  &    0.093 $\pm$ 0.031    \\
108& Aug.9   & 29.61646  &    0.082 $\pm$ 0.031    \\
109& Aug.9   & 29.67886  &    0.050 $\pm$ 0.026    \\
110& Aug.9   & 29.74029  &    0.070 $\pm$ 0.026    \\
111& Aug.9   & 29.80190  &    0.037 $\pm$ 0.025    \\
112& Aug.9   & 29.86328  &   -0.043 $\pm$ 0.026    \\
113& Aug.9   & 29.92464  &    0.034 $\pm$ 0.026    \\
114& Aug.9   & 29.98671  &   -0.030 $\pm$ 0.025    \\
115& Aug.9   & 30.04833  &   -0.036 $\pm$ 0.023    \\
116& Aug.9   & 30.10971  &   -0.137 $\pm$ 0.026    \\
117& Aug.9   & 30.23262  &   -0.075 $\pm$ 0.025    \\
118& Aug.9   & 30.30721  &   -0.120 $\pm$ 0.026    \\
119& Aug.9   & 30.36858  &   -0.075 $\pm$ 0.025    \\
120& Aug.9   & 30.43001  &   -0.165 $\pm$ 0.025    \\
121& Aug.9   & 30.49167  &   -0.107 $\pm$ 0.028    \\
122& Aug.9   & 30.55305  &   -0.100 $\pm$ 0.031    \\
123& Aug.9   & 30.63248  &   -0.167 $\pm$ 0.025    \\
124& Aug.9   & 30.69387  &   -0.162 $\pm$ 0.024    \\
125& Aug.9   & 30.75544  &   -0.136 $\pm$ 0.025    \\
126& Aug.9   & 30.81677  &   -0.158 $\pm$ 0.025    \\
127& Aug.9   & 30.87809  &   -0.151 $\pm$ 0.024    \\
128& Aug.9   & 31.05442  &   -0.291 $\pm$ 0.025    \\
129& Aug.9   & 31.42471  &   -0.230 $\pm$ 0.026    \\
130& Aug.9   & 32.79421  &   -0.061 $\pm$ 0.028    \\
131& Aug.9   & 33.07137  &   -0.026 $\pm$ 0.030    \\
132& Aug.12  & 99.73628  &   -0.390 $\pm$ 0.050    \\
133& Aug.12  & 99.79843  &   -0.332 $\pm$ 0.053    \\
134& Aug.12  & 99.88173  &   -0.238 $\pm$ 0.055    \\
135& Aug.12  & 99.94330  &   -0.282 $\pm$ 0.052    \\
136& Aug.12  & 100.00467 &	-0.270 $\pm$ 0.045    \\
137& Aug.12  & 100.06596 &	-0.211 $\pm$ 0.042    \\
138& Aug.12  & 100.12734 &	-0.339 $\pm$ 0.048    \\
139& Aug.12  & 100.33770 &	-0.225 $\pm$ 0.034    \\
140& Aug.12  & 100.39931 &	-0.167 $\pm$ 0.033    \\
141& Aug.12  & 100.46064 &	-0.249 $\pm$ 0.043    \\
142& Aug.12  & 100.52200 &	-0.250 $\pm$ 0.040    \\
143& Aug.12  & 100.58336 &	-0.200 $\pm$ 0.037    \\
144& Aug.12  & 100.64521 &	-0.156 $\pm$ 0.039    \\
145& Aug.12  & 100.70649 &	-0.392 $\pm$ 0.058    \\
146& Aug.12  & 100.76810 &	-0.231 $\pm$ 0.033    \\
147& Aug.12  & 100.82973 &	-0.240 $\pm$ 0.031    \\
148& Aug.12  & 100.89125 &	-0.218 $\pm$ 0.034    \\
149& Aug.12  & 100.95269 &	-0.175 $\pm$ 0.029    \\
150& Aug.12  & 101.01404 &	-0.119 $\pm$ 0.036    \\
151& Aug.12  & 101.07562 &	-0.122 $\pm$ 0.040    \\
152& Aug.12  & 101.13722 &	-0.153 $\pm$ 0.038    \\
153& Aug.12  & 101.19859 &	-0.111 $\pm$ 0.032    \\
154& Aug.12  & 101.26019 &	-0.172 $\pm$ 0.031    \\
155& Aug.12  & 101.32154 &	-0.146 $\pm$ 0.034    \\
156& Aug.12  & 101.38279 &	-0.109 $\pm$ 0.038    \\
157& Aug.12  & 101.44414 &	-0.170 $\pm$ 0.037    \\
158& Aug.12  & 101.50550 &	-0.059 $\pm$ 0.028    \\
159& Aug.12  & 101.56772 &	-0.109 $\pm$ 0.025    \\
160& Aug.12  & 101.62931 &	-0.048 $\pm$ 0.027    \\
161& Aug.12  & 101.75224 &	0.013  $\pm$ 0.031    \\
162& Aug.12  & 101.81387 &	0.008  $\pm$ 0.032    \\
163& Aug.12  & 101.93697 &	-0.183 $\pm$ 0.025    \\
164& Aug.12  & 101.99857 &	0.024  $\pm$ 0.028    \\
165& Aug.12  & 102.06017 &	0.021  $\pm$ 0.030    \\
166& Aug.12  & 102.12178 &	0.053  $\pm$ 0.036    \\
167& Aug.12  & 102.18341 &	-0.014 $\pm$ 0.034    \\
168& Aug.12  & 102.24477 &	-0.196 $\pm$ 0.037    \\
169& Aug.12  & 102.30640 &	-0.012 $\pm$ 0.034    \\
170& Aug.12  & 102.36801 &	0.086  $\pm$ 0.036    \\
171& Aug.12  & 102.42936 &	0.150  $\pm$ 0.032    \\
172& Aug.12  & 102.49235 &	0.133  $\pm$ 0.035    \\
173& Aug.12  & 102.55367 &	0.141  $\pm$ 0.032    \\
174& Aug.12  & 102.61504 &	0.046  $\pm$ 0.035    \\
175& Aug.12  & 102.67640 &	0.028  $\pm$ 0.034    \\
176& Aug.12  & 102.92412 &	0.170  $\pm$ 0.047    \\
177& Aug.12  & 103.21338 &	0.202  $\pm$ 0.039    \\
178& Aug.12  & 103.49812 &	0.312  $\pm$ 0.044    \\
179& Aug.12  & 104.28804 &	0.198  $\pm$ 0.037    \\
180& Aug.12  & 104.35062 &	0.202  $\pm$ 0.045    \\
181& Aug.12  & 104.41201 &	0.325  $\pm$ 0.052    \\
182& Aug.12  & 104.47354 &	0.248  $\pm$ 0.046    \\
183& Aug.12  & 104.53492 &	0.186  $\pm$ 0.039    \\
184& Aug.12  & 104.59631 &	-0.045 $\pm$ 0.034    \\
185& Aug.12  & 104.65873 &	0.267  $\pm$ 0.044    \\
186& Aug.12  & 104.72009 &	0.254  $\pm$ 0.045    \\
187& Aug.12  & 104.90471 &	0.137  $\pm$ 0.036    \\
188& Aug.12  & 104.99241 &	0.339  $\pm$ 0.076    \\
189& Aug.12  & 105.05683 &	0.156  $\pm$ 0.047    \\
190& Aug.12  & 105.11857 &	0.094  $\pm$ 0.073    \\
191& Aug.12  & 105.18058 &	0.123  $\pm$ 0.039    \\
192& Aug.12  & 105.30671 &	0.013  $\pm$ 0.065    \\
193& Aug.12  & 105.36855 &	0.302  $\pm$ 0.062    \\
194& Aug.12  & 105.42993 &	0.235  $\pm$ 0.045    \\
195& Aug.12  & 105.49139 &	0.168  $\pm$ 0.036    \\
196& Aug.12  & 105.67662 &	0.218  $\pm$ 0.050    \\
197& Aug.12  & 105.73822 &	0.217  $\pm$ 0.051    \\
198& Aug.12  & 105.79961 &	0.122  $\pm$ 0.054    \\
199& Aug.12  & 105.86091 &	-0.106 $\pm$ 0.053    \\
200& Aug.12  & 105.92229 &	-0.042 $\pm$ 0.056    \\
201& Aug.12  & 105.98367 &	0.034  $\pm$ 0.043    \\
202& Aug.12  & 106.04529 &	0.129  $\pm$ 0.049    \\
203& Aug.12  & 106.10664 &	0.258  $\pm$ 0.053    \\
204& Aug.12  & 106.16805 &	0.111  $\pm$ 0.046    \\
205& Aug.12  & 106.22961 &	0.116  $\pm$ 0.055    \\
%\sidehead{I'm a side head:}
\enddata
\tablenotetext{a}{Time since UT 2013 August 8.00000. The middle of integration times is taken.}
\tablenotetext{b}{Red magnitude relative to field stars in background.}
%% You can append references to a table using the \tablerefs command.
%\tablerefs{
%(1) Barbuy, Spite, \& Spite 1985; (2) Bond 1980; (3) Carbon et al. 1987}
\end{deluxetable}

 % table 3
% label lightphot
\clearpage

\begin{deluxetable}{llrrc}
\tabletypesize{\scriptsize}
\tablewidth{0pt}
\tablecaption{Measured Colors of 2003EH1 and Small Body Populations \label{distcolor}}
\tablehead{\colhead{Object} & \colhead{$V - R$} & \colhead{$S'$} & \colhead{$N$$^{a}$} &\colhead{Source} }         
\startdata   
2003EH1$^{b}$   &  0.39$\pm$0.01    &   3.7$\pm$0.9   &  -   &(1) \\ 
KBOs$^{c}$      &  0.59$\pm$0.12    &  22.0$\pm$10.9   & 297  &(2) \\
Centaurs        &  0.54$\pm$0.01    &  17.5$\pm$0.9   &  32  &(2) \\ 
Damocloids$^{d}$&  0.48$\pm$0.01    &   12.0$\pm$0.9  &  12  &(3) \\
Nuclei$^{e}$    &  0.45$\pm$0.02    &   9.2$\pm$1.8   &  12  &(4) \\
Dead Comets     &  0.44$\pm$0.02   &    8.3$\pm$1.8  &   12  &(4) \\
Trojans$^{f}$   &  0.46$\pm$0.01    &   10.1$\pm$0.9  & 451  &(5) \\
D-types         &  0.45$\pm$0.01   &     9.2$\pm$0.9  &  19  &(6) \\
MBCs$^{g}$      &  0.37$\pm$0.01   &    1.8$\pm$0.9   &   6  &(7) \\
Solar Color     &  0.35$\pm$0.01   &    0.0$\pm$0.9   &  -   &(8) \\
%\cutinhead{This is a cut-in head}
%\sidehead{I'm a side head:}
\enddata
\tablecomments{$S'$ from the relation, ${\it (V - R)}$ = ${\it (V - R)_{\odot}}$+2.5\,log$[(2+S'\Delta\lambda)/(2-S'\Delta\lambda)]$, 
where ${\it (V- R)}$ and ${\it (V-R)_{\odot}}$=0.35 are the colors of the object and the Sun respectively, 
and $\Delta \lambda$=1000\,$\rm \AA$ is the difference between the ${\it V}$- and ${\it R}$- filters~\citep{LuuJewitt1990}.}
\tablenotetext{a}{Number of objects in the population.}
\tablenotetext{b}{The weighted mean of measurements from Table~\ref{colorEH1}.  }
\tablenotetext{c}{Kuiper Belt Objects.}
\tablenotetext{d}{Inactive cometary nuclei of Halley-family and long-period comets with $T_J \le 2$}
\tablenotetext{e}{Cometary Nuclei.}
\tablenotetext{f}{Jupiter Trojans.}
\tablenotetext{g}{Main belt comets.}
%% You can append references to a table using the \tablerefs command.
\tablerefs{(1) This work, (2) ~\cite{Pex15}; (\cite{JL01} and \cite{Bau03} are included), (3) ~\cite{J05}, 
(4) ~\cite{Jewitt2002}, (5) ~\cite{Szab07}; (see also \cite{JewittLuu1990} and \cite{Karlsson2009}), 
(6) ~\cite{Fi94}, (7)  ~\cite{Hsieh2004,Hsieh2009,Hsieh2010,Hsieh2011,Hsieh2013,Hsieh2015,Jewitt2009}, (8)~\cite{H06}}
\end{deluxetable}

% table 4
% label distcolor
\clearpage

%%%%%%%%%%%%%%%%%%%%%%%%%%%%%%%%%%%%%%%%%%%%%%
%% Figures

% Fig 1
% frame 40 " x 25 "
\clearpage

% Fig 2
\clearpage

% Fig 3
\clearpage

% Fig 4
\clearpage

% Fig 5
\clearpage

%% The equation environment wil produce a numbered display equation.

%% Included in this acknowledgments section are examples of the
%% AASTeX hypertext markup commands. Use \url without the optional [HREF]
%% argument when you want to print the url directly in the text. Otherwise,
%% use either \url or \anchor, with the HREF as the first argument and the
%% text to be printed in the second.
\clearpage
\newpage


%% in parentheses, after the keyword, but they will not be verified.

%{\it Facilities:} \facility{Nickel}, \facility{HST (STIS)}, \facility{CXO (ASIS)}.



\end{document}